\documentstyle[aps,multicol,epsfig,prb]{revtex}
\begin{document}
\draft
\title{Failure Probabilities and Tough-Brittle Crossover of Heterogeneous Materials with Continuous Disorder}
\author{B. Q. Wu, and P. L. Leath}
\address{Department of Physics and Astronomy, Rutgers University, 136 Frelinghuysen Road, Piscataway, New Jersey 08854-8019}

\maketitle

\widetext
\begin{abstract}
The failure probabilities or the strength distributions of heterogeneous 1D systems with continuous local strength distribution and local load sharing have been studied using a simple, exact, recursive method. The fracture behavior depends on the local bond-strength distribution, the system size, and the applied stress, and crossovers occur as system size or stress changes. In the brittle region, systems with continuous disorders have a failure probability of the modified-Gumbel form, similar to that for systems with percolation disorder. The modified-Gumbel form is of special significance in weak-stress situations. This new recursive method has also been generalized to calculate exactly the failure probabilities under various boundary conditions, thereby illustrating the important effect of surfaces in the fracture process.
\end{abstract}
\pacs{PACS numbers: 46.30.Nz, 62.20.Mk, 81.40.Np, 83.50.Tq}

\begin{multicols}{2}
\narrowtext
\section{Introduction}

Due to their importance in many scientific and engineering fields, breakdown phenomena in heterogeneous materials have been studied extensively in recent years~\cite{hr}. It has been well established that the tensile strength of heterogeneous materials is dominated by the "weakest" regions of the systems, or by the extreme fluctuations of the stress fields. Thus the fracture properties of a given sample are highly non-intrinsic, non-self-averaging, and strongly depend on the specific realization of the strengths of individual bonds based on the heterogeneity and microstructure. This dependence of fracture on extreme statistics, hence the lack of self-averaging, often makes mean-field theories rather dubious. The failure probability $F_n(\sigma)$ of a heterogeneous system of size $n$ under external tensile stress $\sigma$ or identically the cumulative distribution function (cdf) of its strength can be described by one of the distribution forms of extreme value statistics, such as the Weibull distribution
\begin{equation}
F_n(\sigma)=1-\exp[-n(\sigma /\sigma_0)^m] \ , \label{wei}
\end{equation} 
or the {\em modified-Gumbel} distribution~\cite{dlb}
\begin{equation}
F_n(\sigma)=1-\exp[-n\exp(-\Lambda/\sigma^{1/\psi})]  \ . \label{gum}
\end{equation} 
The modified-Gumbel form was introduced by Duxbury {\em et al}~\cite{dlb} in their study of percolation disorder on diluted 2D fuse networks (square lattice) in which identical lattice bonds (same conductance and breakdown threshold) are present with a probability $p$. They observed that in general the failure of the first few bonds in the diluted network worsens the situation fatally and leads to the failure of the entire system, or brittle fracture. With this result and those of numerical simulations in tough and brittle systems~\cite{kbrah} it was conjectured that the failure probability is described for tough fracture by Eq.~\ref{wei}, and for brittle fracture, by Eq.~\ref{gum}. Statistically the degree of brittleness of fracture can be characterized by the ratio of the size of typical local damage that triggers the system failure to the system size. That is, a fracture is said to be brittle if this ratio is small and tough if it is of order one. Thus the degree of brittleness of fracture depends on the disorder type (modeled by the distribution of the bond strengths) and the system size~\cite{kbrah,cl}. But also it depends on the applied stress since the critical damage shows a dependence on the external stress which we shall discuss below. A major difference between percolation disorder and continuous disorder is that for percolation disorder the failure stress of the system $\sigma_b$ is approximately the fracture stress $\sigma_1$ for the first bond to break, while for the continuous disorder one may have $\sigma_b \gg \sigma_1 \rightarrow \inf\{\sigma: G(\sigma)>0\}$, where $G(\sigma)$ is the local bond-strength distribution. 

The fracture of materials is typically highly localized in the cracks that nucleate and grow due to the local stress enhancement at crack tips. In real materials, because of the long-range elastic interaction, the stress redistribution follows a power law, $\Delta\sigma\sim r^{-g}$, where $\Delta\sigma$ is the stress increase on a bond at distance $r$ from a crack tip. The extreme cases of this picture lead to two important models which because of their simplicity have received considerable attention. First, for $g \rightarrow 0$ one obtains the {\em global-load-sharing} or {\em equal-load-sharing} (ELS) model, by which the stress released by cracks is equally shared by all the remaining bonds across the sample. This model, analytically solvable, is mean field-like, and consequently the system strength has been found to be a Gaussian distribution~\cite{s} rather than of a type of extreme statistics. Second, for $g \rightarrow \infty$ one obtains the {\em local-load-sharing} (LLS) model, by which the stress released by a crack is shared only by the intact bonds at the crack tips. This model is more realistic in the sense that it does show extreme-statistical aspects. In 1D, an exact, numerical solution of this model is available~\cite{hp,dl} and it has been used in the study of simple 2D systems with linear cracks which can be approximately treated as a stack of independent 1D systems~\cite{dl}. Here we propose a simple, exact, recursive method which may lead to an ultimate analytical solution to the 1D problem, and with its help, we obtain some insight into the form of the strength distribution in more general cases. In addition, by using this powerful method, we have re-derived the strength distribution of the diluted model analytically.

Consider an array of $n$ bonds, each of which is assigned a random strength threshold drawn from a continuous distribution function $G(x)$. In this article we shall use, for example, the Uniform(0,1) distribution: $G(x)=x$, $0\le x\le 1$ and the Weibull(m) distribution: $G(x)=1-\exp (-x^m)$. We assume the disorder is quenched and when the stress on a local bond exceeds its assigned strength threshold, it fractures irreversibly and quasi-statically. As an external tensile stress $\sigma$ is applied, some of the bonds may fracture immediately and the stress released will be redistributed among the temporarily surviving bonds and this stress enhancement might trigger a secondary wave of bond fractures, etc. The system fails in one dimension when the last surviving bond fails, or in higher dimensions when a spanning crack forms. It's easy to see that the failure of a composite depends sensitively on how the stress is redistributed among the surviving bonds. In the {\em local-load-sharing} (LLS) model, the stress released by the formation of an internal crack is equally shared by the two bonds at the tips. However, if the crack is on the end of the sample, the simplest method is to assume that there is an intact bond outside to bear the stress. This is called an {\em interior} boundary condition, and is strictly applicable only when the sample is embedded in a larger one with intact bonds. Thus if a bond has $k$ neighboring failed bonds in the LLS model, the stress on this bond is $\sigma_k=(1+k/2)\sigma$, where $\sigma$ is the applied stress, and this bond survives with probability $W_k(\sigma)\equiv 1-G(\sigma_k)$. We shall discuss this and a number of more complicated and realistic boundary conditions, namely, the periodic, semi-open, and open boundary conditions~\cite{cl}.

\section{Recursive Solution of the 1D-Lattice Model with LLS}

In this section, we discuss the recursive solution of the 1D LLS model with interior boundary condition. In order to find the failure probability (or strength distribution) of the composite given the above, we develop a powerful and simple recursive method, simpler than the one used previously~\cite{dl}. To find the failure probability, we need to evaluate the sum of the survival probabilities for all configurations (except the one with all failed bonds) and we define $S_{n,l}$ to be the sum of the {\em survival} probabilities for all possible configurations in a sample of size $n$ {\em with} $l$ {\em fractured bonds on the far right end}, so that the failure probability can be written $F_n(\sigma)=1-\sum_{l=0}^{n-1}S_{n,l}(\sigma)$. To find the recursive relation for $S_{n,l}$, we consider the following configuration
\begin{figure}
  \begin{center}
    \epsfig{file=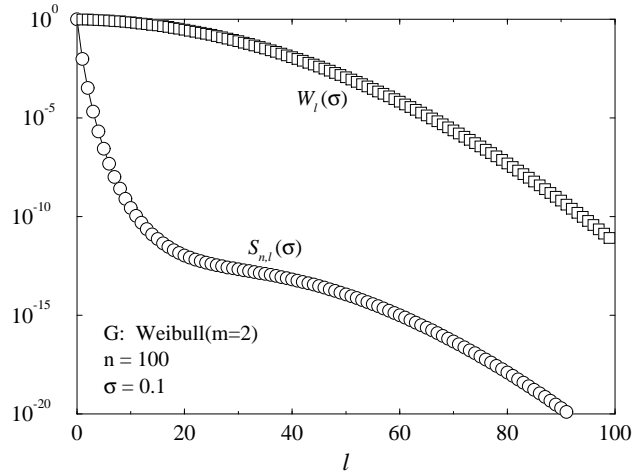,width=6.2cm,angle=-90,clip=1}
  \end{center}
\caption{$S_{n,l}(\sigma)$ as a function of $l$, the number of broken bonds on the right end. Also shown is $W_l(\sigma)$ vs. $l$, where $W_l(\sigma)=1-G[(1+l/2)\sigma]$. Bond-strength distribution $G$: Weibull(2); system size: $n=100$; applied stress $\sigma=0.1$.} 
\end{figure} 
$$\underbrace{\cdots\cdots 1\overbrace{0\cdots 0}^r1\overbrace{0\cdots 0}^l}_n $$
where "1" represents an intact bond and "0" represents a fractured bond. Suppose we remove the intact bond and the $l$ broken bonds at the rightmost end, we are left with a similar configuration, but of smaller size $n-l-1$, therefore the survival probabilities $\{S_{n,l}\}$ should satisfy the recursive relation
\begin{equation}
S_{n,l}=F_{n-l-1}W_{n-1}F_l+\sum_{r=0}^{n-l-2}S_{n-l-1,r}W_{r+l}F_l \ ,
\end{equation}
where as a special case the first term corresponds to the surviving configuration having only one intact bond. We further define $S_{k,k}$ to be the {\em failure} probability of a such system of size $k$, {\em i.e.} $S_{k,k}\equiv F_k$, then the above results can be rewritten in the compact form
\begin{equation}
\left\{ \begin{array}{l}
            S_{n,n}=1-{\displaystyle \sum_{l=0}^{n-1}S_{n,l}}\\
            S_{n,l}={\displaystyle \sum_{r=0}^{n-l-1}S_{n-l-1,r}W_{r+l}S_{l,l}} \ , \qquad l=0,\cdots , n-1 \ ,
          \end{array} \right. \label{com}
\end{equation}
with an initial condition $S_{0,0}=1$ which serves as a "seed" for the entire calculations. This exact recursion formula can be evaluated numerically very efficiently. Despite its simplicity, so far we have found that it's hard to analytically solve Eq.~\ref{com}. Nevertheless, by noticing that both $W_l$ and $S_{n,l}$ decay rapidly with increasing $l$, and that for $l$ small, $S_{n,l}/S_{n-1,l}$ is essentially a constant, we find an excellent approximation to Eq.~\ref{com}. Fig.~1 shows the $l$ dependence of $S_{n,l}$ and $W_l$ for a given system of size $n=100$ with bond strength distribution Weibull(2) under stress $\sigma=0.1$; it is obvious that these functions decay in some exponential manner. Fig.~2 shows the ratio of survival probabilities of two systems of sizes $n$ and $n-1$ respectively with $l$ broken bonds on the right end, $S_{n,l}/S_{n-1,l}$, as a function of $l$, under the same conditions. The ratio is very close to $1$ and nearly a constant. 
\begin{figure}
  \begin{center}
    \epsfig{file=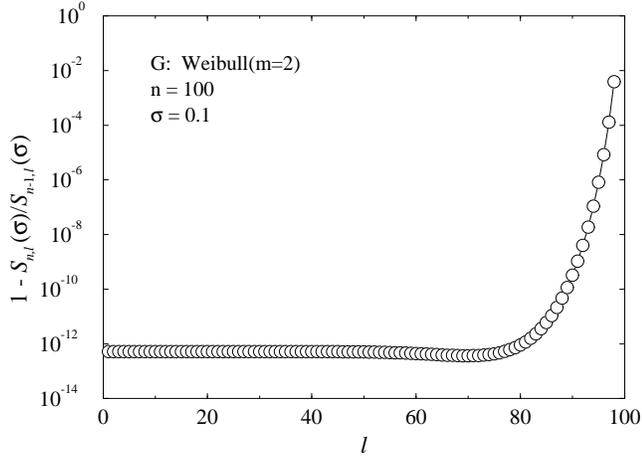,width=6.1cm,angle=-90,clip=1}
  \end{center}
\caption{$1-S_{n,l}(\sigma)/S_{n-1,l}(\sigma)$ as a function of $l$. The survival probability ratio $S_{n,l}(\sigma)/S_{n-1,l}(\sigma)$ is very close to $1$ and nearly independent of $l$. Bond-strength distribution $G$: Weibull(2); system size: $n=100$; applied stress $\sigma=0.1$.}
\end{figure} 

From Eq.~\ref{com} it is obvious that
$$\sum_{r=0}^{n-l-1}S_{n-l-1,r}W_{r+l}=\frac{S_{n,l}}{S_{n-1,l}}\sum_{r=0}^{n-l-2}S_{n-l-2,r}W_{r+l} \ ,$$
and also that
\begin{eqnarray*}
1-S_{n,n}&=&\sum_{l=0}^{n-1}\sum_{r=0}^{n-l-1}S_{n-l-1,r}W_{r+l}S_{l,l}\\
         &=&\sum_{l=0}^{n-2}\frac{S_{n,l}}{S_{n-1,l}}\sum_{r=0}^{n-l-2}S_{n-l-2,r}W_{r+l}S_{l,l}\\
         & &\makebox[3cm]{}+W_{n-1}S_{n-1,n-1}\ .
\end{eqnarray*}
So since $S_{n,l}/S_{n-1,l}$ is nearly independent of $l$, we obtain
\begin{eqnarray*}
1-S_{n,n}&\approx&\frac{S_{n,0}}{S_{n-1,0}}\sum_{l=0}^{n-2}\sum_{r=0}^{n-l-2}S_{n-l-2,r}W_{r+l}S_{l,l} \\
         &&\makebox[3cm]{} +W_{n-1}S_{n-1,n-1} \ ,
\end{eqnarray*}
or
\begin{eqnarray}
\nonumber
1-S_{n,n}&=&\frac{S_{n,0}}{S_{n-1,0}}(1-S_{n-1,n-1}) + W_{n-1}S_{n-1,n-1},\\
         && \makebox[3cm]{}  n=1,2,\cdots \ , \label{app}
\end{eqnarray}
where the second term $W_{n-1}S_{n-1,n-1}$ is the survival probability of the configuration with only one single intact bond at the left end, and the first and dominant term is the total survival probability for all the other configurations. 
\begin{figure}
  \begin{center}
    \epsfig{file=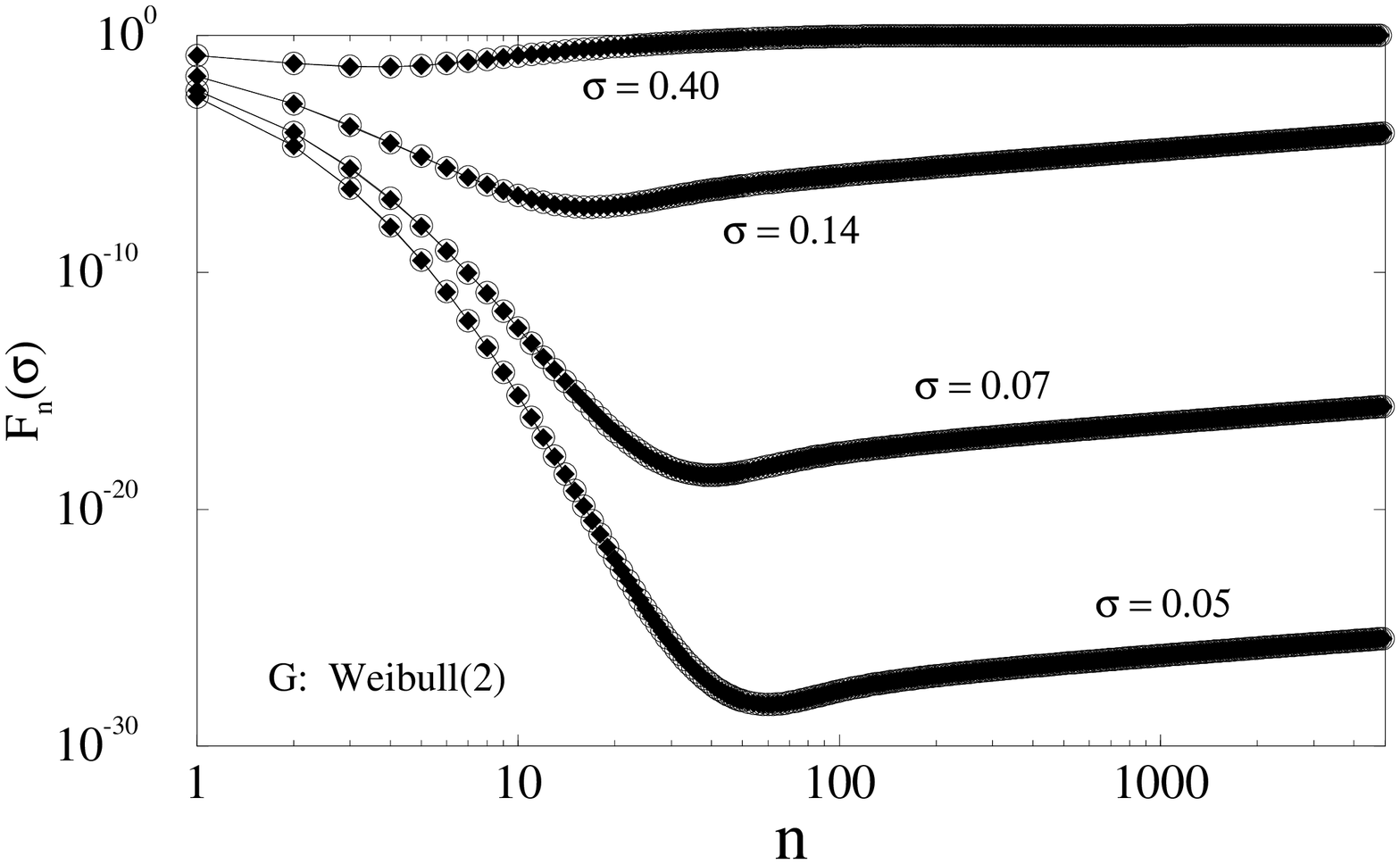,width=5.5cm,angle=-90,clip=}
    \epsfig{file=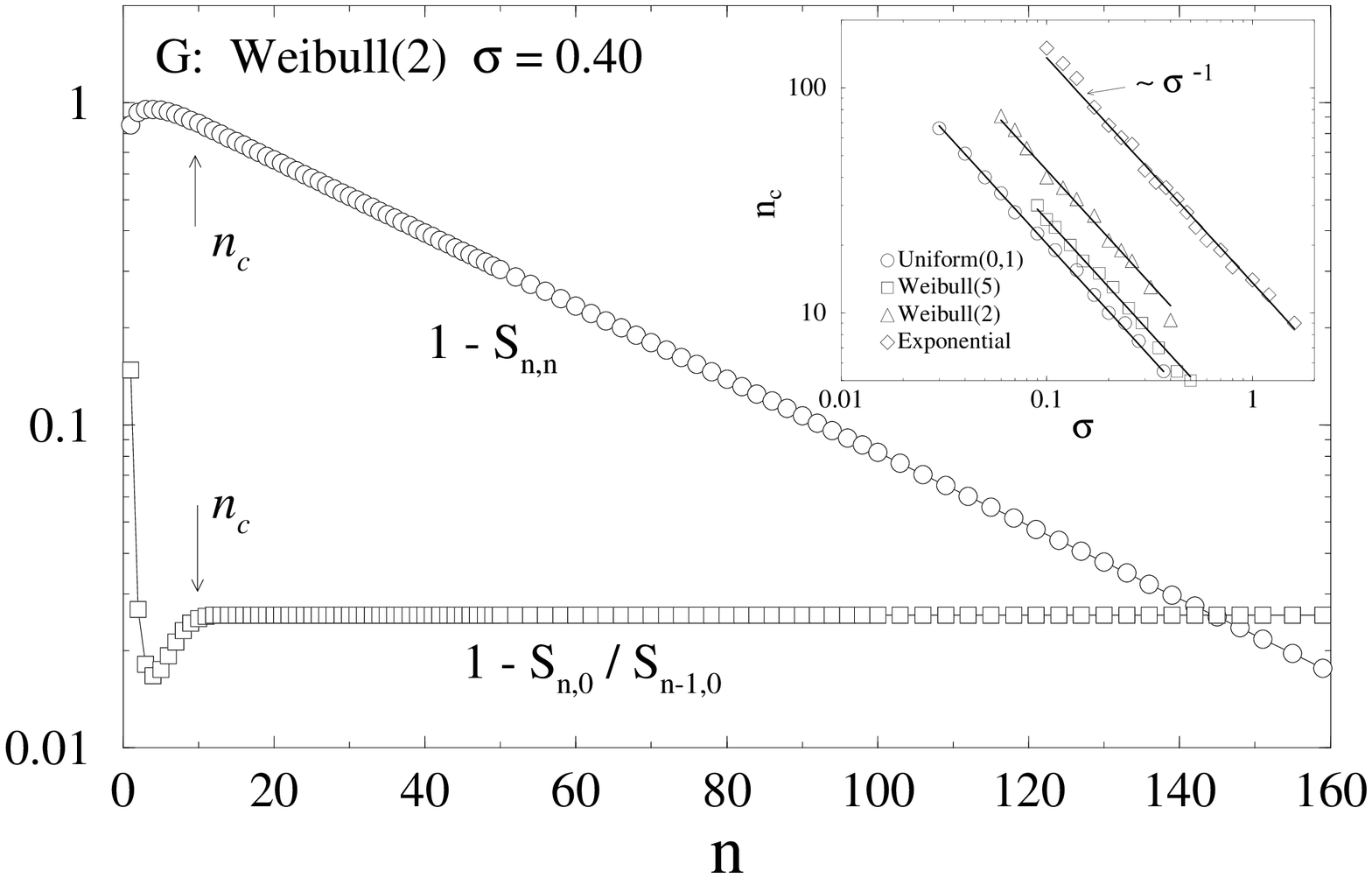,width=5.cm,angle=-90,clip=}
  \end{center}
\caption{(a) Failure probability as a function of system size under different stresses. Open circle: exact numerical data by Eq.~\ref{com}; filled diamond: approximation with Eq.~\ref{app}. (b) The determination of a critical size $n_c$. For $n>n_c$, $1-S_{n,n}(\sigma)$ decays exponentially with $n$, and $1-S_{n,0}(\sigma)/S_{n-1,0}(\sigma)$ becomes essentially a constant. Inset: stress dependence of $n_c$, $n_c\propto \sigma^{-1}$.}
\end{figure}

This is an extremely good approximation, which gives both the large- and small-$n$ behaviors of $S_{n,n}(\sigma)$ or $F_n(\sigma)$ very well, as shown in Fig.~3(a) where the exact results of numerically evaluating Eq.~\ref{com} are compared with Eq.~\ref{app}. We thus focus on the ratio $S_{n,0}/S_{n-1,0}$ which gives the primary behavior of the fracture for sample of size $n$. First, for sufficiently large $n$, we notice that in Fig.~3(b), $S_{n,0}/S_{n-1,0}$ goes to a constant. Thus by neglecting the single-bond term in Eq.~\ref{app}, we obtain that asymptotically the survival probability of a system decays exponentially with $n$, which means that a system is always brittle for $n$ sufficiently large. Also, Eq.~\ref{app} shows its consistency with the weak link condition, with which a large system can be treated as a collection of smaller sub-systems that are weakly related or nearly independent. From Eq.~\ref{app}, it can be easily shown that 
\begin{equation}
(1-S_{i,i})(1-S_{n-i,n-i})\approx (1-S_{j,j})(1-S_{n-j,n-j}) \ ,
\end{equation}
for $i$, $j$, $n-i$, and $n-j$ all much greater than one, given $S_{n,0}/S_{n-1,0}$ goes to a constant. This can be interpreted as that if a large system is divided into parts, then the product of survival probabilities for all parts does not depend on how the system is divided, which is also guaranteed by the weak-link condition. Second, for $n$ small the $n$ dependence of the failure probability has a more complicated behavior as shown in Fig.~3(b) and the system is tough. This suggests the existence of a crossover between different statistics, and allows us to specify a tough-to-brittle size scale $n_c$, such that the breakdown behavior of a system of size larger than $n_c$ becomes brittle. In other words, a system of size $n \gg n_c$ can be treated as a composite of subsystems of size $n_c$, which shows {\em brittle}-like properties for cracks larger than the scale $n_c$. This size scale is analogous to the critical size in the nucleation process of a first-order phase transition. Given a sufficiently large system, the failure of the disordered composite behaves like a first-order phase transition~\cite{zrsv} in that with the nucleation of a critical crack, the system fails. It's reasonable to define $n_c$ to be the system size beyond which $S_{n,0}/S_{n-1,0}$ becomes essentially a constant [Fig.~3(b)]. The numerical results show that $n_c \propto \sigma^{-1}$ independent of the local bond threshold distribution $G(x)$, but the coefficient does depend on $G(x)$, as shown in Fig.~3(b) inset.

We thus come to a two-scale picture of fracture: locally within size $n_c$ the fracture is tough and globally the fracture is brittle at scale $n_c$. This confirms the picture proposed by Curtin~\cite{c} recently. In his letter, Curtin studied planar systems under stresses in the normal direction and observed the existence of a critical size $n_c$, according to which the system could be characterized as a collection of subsystems of size $n_c$. A subsystem in Curtin's picture was formulated in an equal load sharing (ELS) model, while the whole system became brittle in that failure of any single subsystem necessarily led to the breakdown of the whole system. The question remaining is to determine $n_c$. In our numerically solvable 1D model, we find a power-law stress dependence of $n_c\propto \sigma^{-1}$ when $\sigma$ is small. As we'll describe below, with $n_c \propto \sigma^{-\alpha}$, the failure probability belongs to the "modified-Gumbel" family of the extreme statistics.

For our method we find, and for more general models suggest, that for a system of fixed size $n$ with applied stress $\sigma$, the failure distribution will generally have these behavioral regions as follows:

a) $n < n_c$ : This is the {\em tough} region, which corresponds to the initial downhill portion of the curves in Fig.~3(a). Generally this region is characterized by very small stresses (which may be the normal operating stresses in practical applications) and smaller system sizes. The expansion of Eq.~\ref{com} gives the exact failure probabilities and, in particular it is applicable for very small $n$'s. The failure probability is a superposition of a huge number of local distributions $G(x)$, and because the failure of the composite is path (or ordering) dependent, the upper bound of the number can be of order $2^{n-1}n!$. Fortunately, we are only interested in the left tail of the failure distribution, which is essential in the extreme (minimum) statistics analysis. Consider a Weibull($m$) local bond-strength distribution: $G(x) = 1 - \exp(-x^m)$, the survival probability of a bond with $k$ fractured neighbor bonds is $W_k(\sigma)=\exp\{-[(1+k/2)\sigma]^m\}$. For the left tail of the distribution $S_{n,n}(\sigma)$ where both $\sigma$ and $n$ are small, we can make Taylor series expansions of $W_k(\sigma)$ about $\sigma =0$ in Eq.~\ref{com}, and we find the failure probabilities for samples of different sizes are given by
\begin{eqnarray*}
&&S_{1,1}=\sigma^m+O(\sigma^{2m}) \ ,\\
&&S_{2,2}=[-1+2\cdot (\frac{3}{2})^m]\sigma^{2m}+O(\sigma^{3m}) \ ,\\
&&S_{3,3}=[1-2\cdot (\frac{3}{2})^m-2^m+4\cdot 3^m-(\frac{3}{2})^{2m}]\sigma^{3m} \\ 
         && \makebox[1.1cm]{}+O(\sigma^{4m}) \ ,\\
&&S_{4,4}=[-1+2\cdot (\frac{3}{2})^m+(\frac{3}{2})^{2m}+2^{2m}+2^{1+m}\\
         &&\makebox[1.1cm]{}-2^{1+m}\cdot 3^m-2\cdot 3^{1+m}-2^{1-3m}\cdot 3^{2m}\cdot 5^m\\
         &&\makebox[1.1cm]{} -2\cdot 5^m+2^{3-m}\cdot 15^m]\sigma^{4m}+O(\sigma^{5m}) \ , \\
&&\cdots \cdots \ .  
\end{eqnarray*}
For weak stresses, we neglect the higher order terms and obtain
\begin{eqnarray}
\nonumber
F_n(\sigma)=S_{n,n}(\sigma)&\approx& c(n,m)\sigma^{mn}\\
    &\approx& 1-\exp(-c(n,m)\sigma^{mn}) \ ,
\end{eqnarray}
where $c(n,m)$ is a coefficient. It is a Weibull distribution with the modulus proportional to the system size $n$. Fig.~4(a) demonstrates the consistency between the numerical results and this analytical formula. All lower-order terms cancel out in the expansion; physically, this can be understood with the fact that the failure of the composite requires the fracture of all the bonds, and each bond has a contribution of $\sim \sigma^m$ to the failure probability. From this point of view, the failure probability for the tough region at the low-stress limit is always of this form, and the load-sharing rule can only affect the coefficient. The determination of the coefficient $c(n,m)$ is hard, but a crude estimate can be made from the $n!$ ways of fracture ordering of the $n$ bonds. For small $n$, due to the LLS, we estimate $c(n,m)\sim (n!)^{\gamma m}$, where $0<\gamma <1$ is a parameter depending upon $m$. Thus the failure probability for the tough region is of the Weibull form
\begin{equation} F_n(\sigma) \sim 1-\exp(-(n!)^{\gamma m} \sigma^{mn}) \ . \end{equation}
The optimum sample size, corresponding to the minimum failure probability as shown in Fig.~1(a), is now easily estimated. By setting $F_{n-1} = F_n$, we obtain the optimal size 
\begin{equation} n_{min} \sim \sigma^{-1/\gamma} \ , \end{equation}
and the corresponding minimum failure probability
\begin{equation} F_{min} \sim \exp(-c \,\sigma^{-1/\gamma}) \ . \end{equation}
\begin{figure}
  \begin{center}
    \epsfig{file=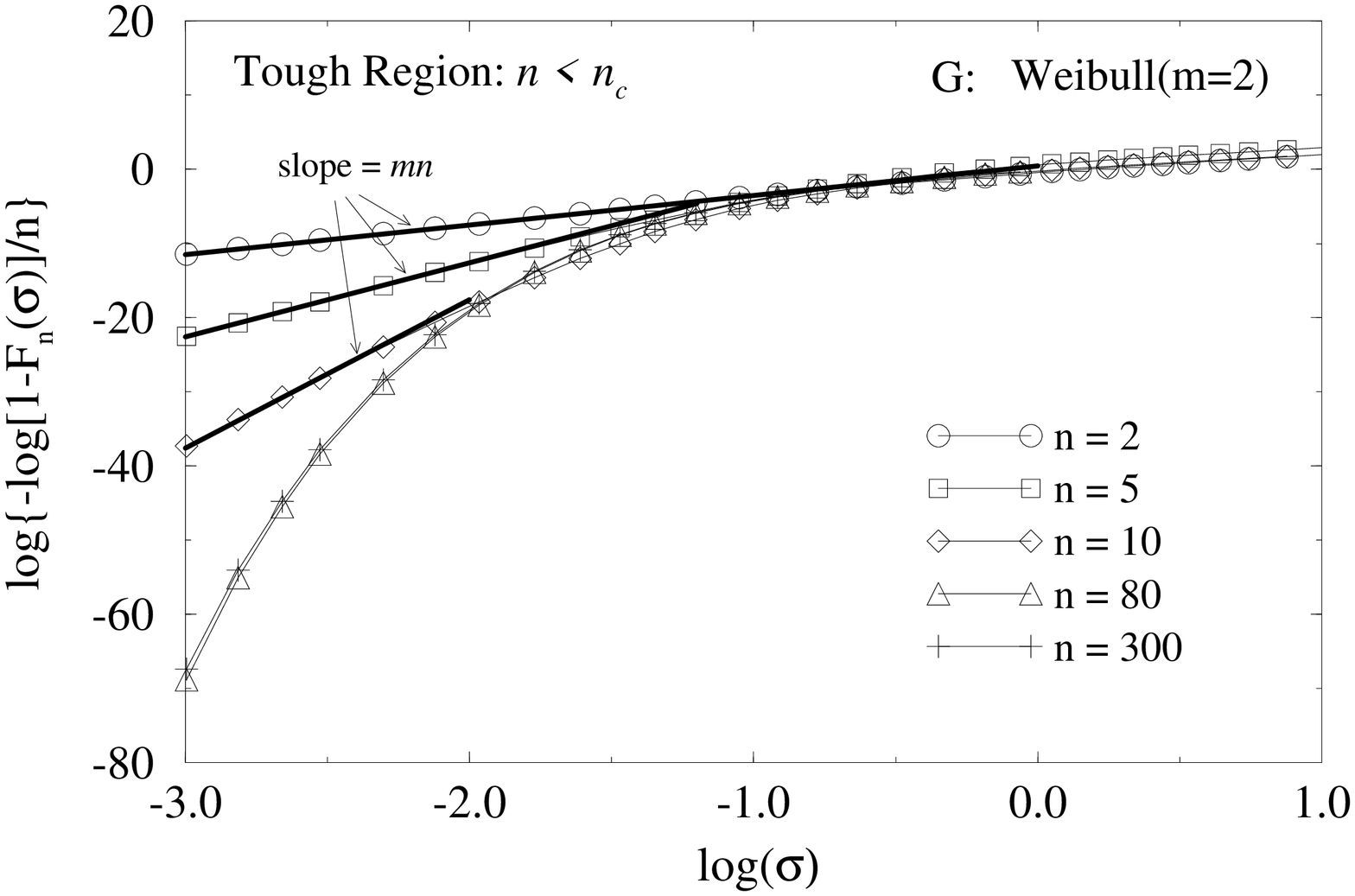,width=5.7cm,angle=-90,clip=}
    \epsfig{file=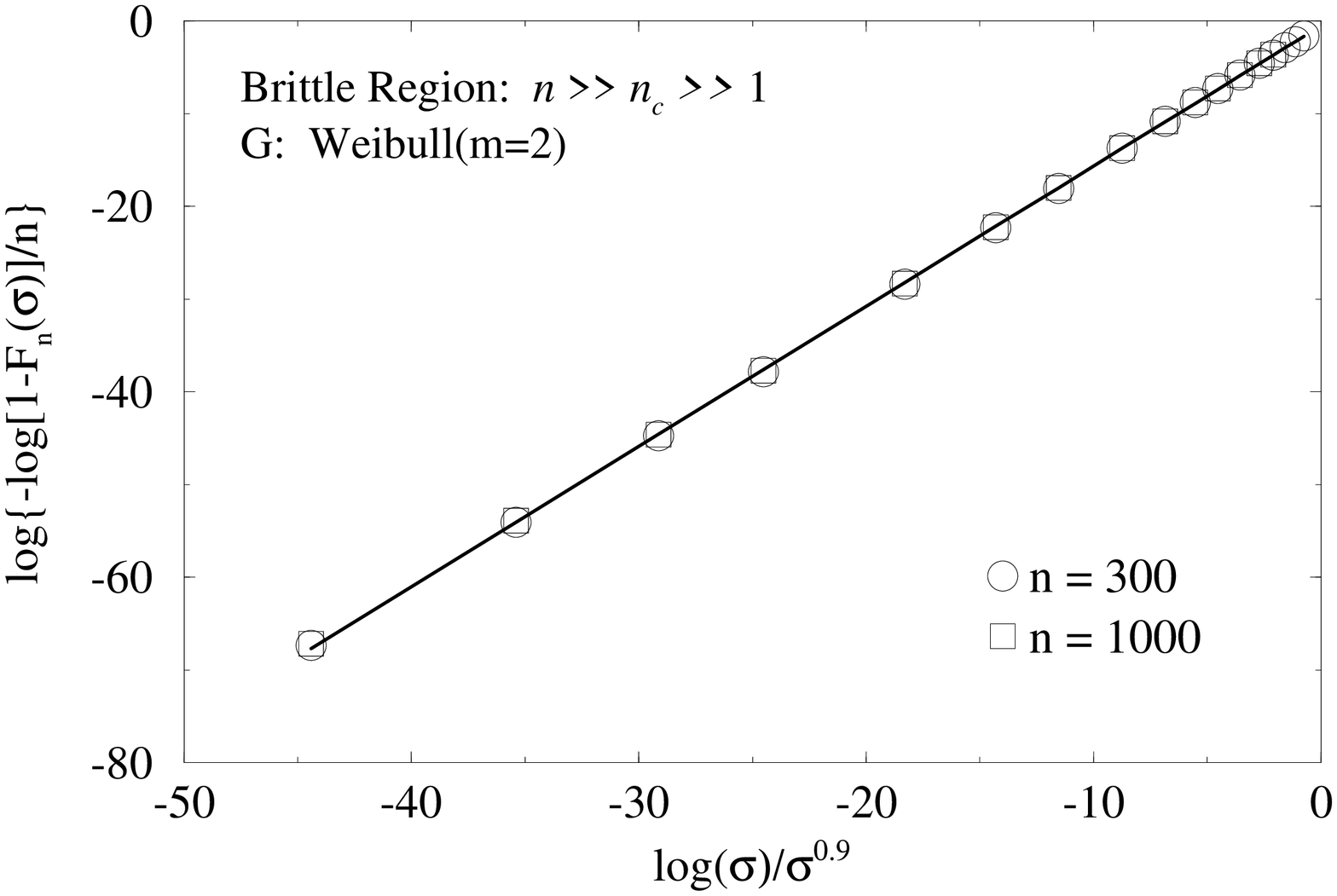,width=5.6cm,angle=-90,clip=}
    \epsfig{file=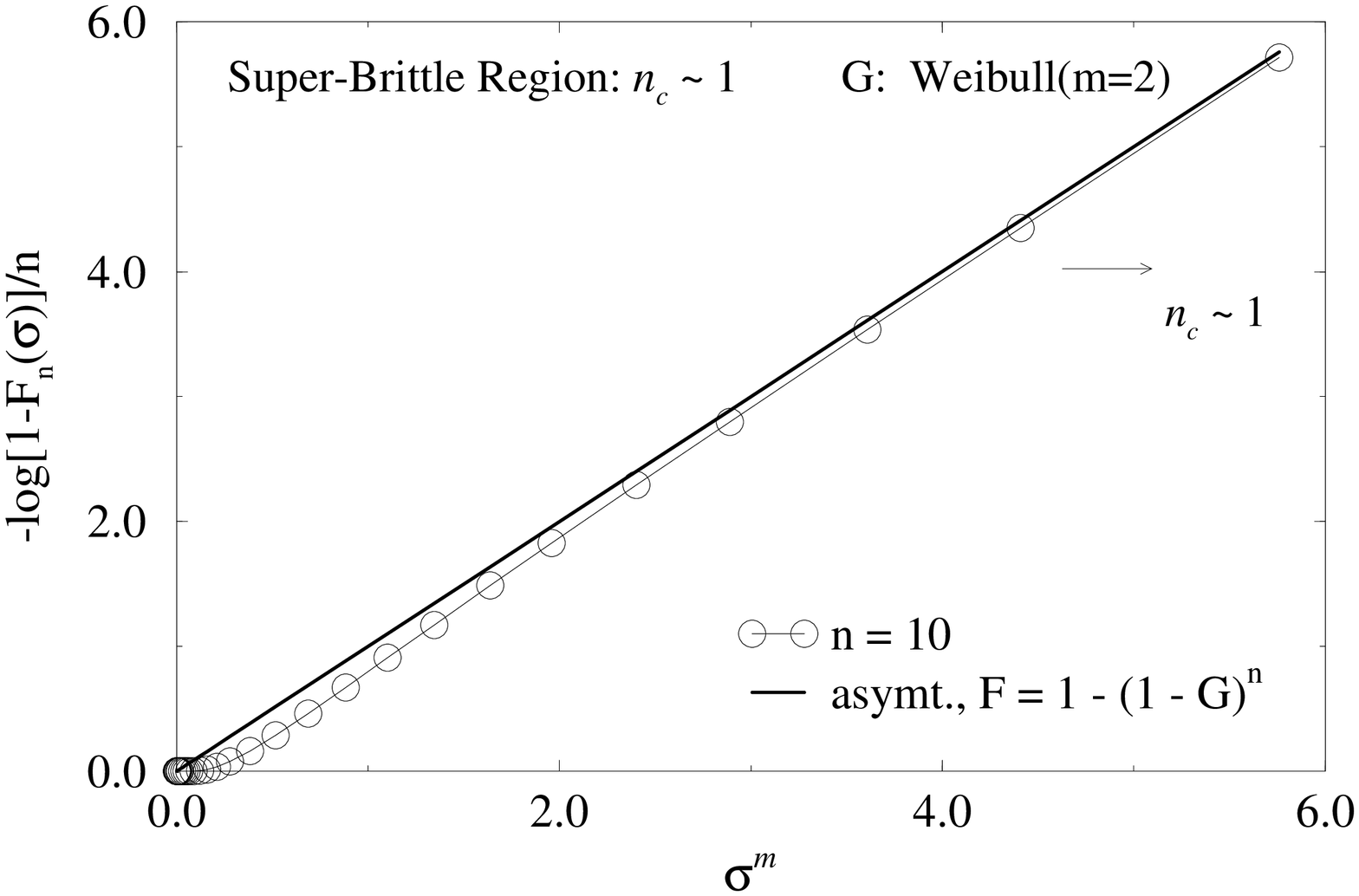,width=5.5cm,angle=-90,clip=}
  \end{center}
\caption {(a) Failure behavior in the tough region, $n<n_c$. Failure probability is a Weibull distribution with modulus $mn$, where $n$ is the system size and $m$ is the Weibull modulus for the local-bond strength distribution. (b) Failure behavior in the brittle region, $n\gg n_c\gg 1$. For a very wide range of failure probability, from $10^{-30}$ to $0.999$, it has a modified-Gumbel form. (c) Failure behavior in the super-brittle region, $n_c\sim 1$. The strength of the system approaches that of the first broken bond.}
\end{figure}
\noindent
For different $G(x)$, we found empirically $\gamma = 0.59$ for Exponential, $0.71$ for Weibull(2), $0.76$ for Weibull(5) and $0.89$ for Uniform(0,1).
 
b) $n \gg n_c \gg 1$ : This is the {\em brittle} region, where the system is macroscopically brittle but microscopically tough. In practice for large system size, most fracture events of real materials fall into this region, since the other regions are compressed. The discussion on the failure probability of a system with percolation disorder under a given stress~\cite{dlb} is still valid in principle for a system with continuous disorder when $n\gg n_c$. Roughly speaking, the failure of the system depends on whether the size of the weakest spot exceeds the critical size $n_c$, and the probability to find a weak spot of size larger than $n_c$ decays exponentially, so we would still expect a "modified-Gumbel" type failure probability for continuous disorder. We observe, as shown in Fig.~3(b), for large $n$ and relatively weak stress, that ${S_{n,0}}/{S_{n-1,0}}$ varies only with $\sigma$, {\em i.e.},
\begin{equation} \frac{S_{n,0}}{S_{n-1,0}}(\sigma) \sim 1-\exp(c\log(\sigma)/\sigma^\alpha) \ , \end{equation}
where exponent $\alpha = 0.90\pm 0.05$ for all local bond distributions tested, is close to $1$, the value expected with the load sharing rule LLS. Combining Eq.~\ref{app}, we get the form of the failure probability
\begin{equation}
F_n(\sigma) = 1-\exp\{-an\exp[c\log(\sigma)/\sigma^\alpha]\} \ , \label{mg}
\end{equation}
where $a$, $c$ and $\alpha$ are parameters to be determined in data fitting. If we use the weak-link hypothesis and treat the system as a collection of subsystems of size $n_c\propto \sigma^{-1}$, then the failure probability can be evaluated by using $F_n \approx 1-(1-F_{n_c})^{n/n_c}$, where $F_{n_c}$ is the failure probability for the tough region at size $n_c$. We thus obtain a similar form to Eq.~\ref{mg}, which implies that the weak-link hypothesis is, in principle, correct. With a minor discrepancy in $\alpha$, Eq.~\ref{mg} belongs essentially to the "modified-Gumbel" family. Thus for continuous disorder, as well as percolation disorder~\cite{dlb}, the failure behavior can still be described with the modified-Gumbel form. Fig.~4(b) demonstrates that the failure probability fits this form extremely well over the very wide range from $10^{-30}$ to $0.999$. 

Although this 1D model is only applicable to problems of breaking a sheet-shaped object such as a piece of paper, we believe and propose that the fracture properties of higher-dimensional materials are similar and the strength-distribution form of Eq.~\ref{mg} is generally applicable. 

c) $n_c \sim 1$ : This is the {\em super-brittle} region of fracture. The stress is so strong that critical nuclei exist almost everywhere, and thus almost all the bonds fail simultaneously. The failure probability is then simply
\begin{equation}
F_n(\sigma) = 1-[1-G(\sigma)]^n \ , \label{real}
\end{equation}
where $G(\sigma)$ is the local bond strength distribution. Fig.~4(c) shows that as stress becomes large, the failure probability approaches form Eq.~\ref{real} asymptotically. It should be noted that the interior boundary condition considered here does not guarantee the total stress to be conserved, especially for small $n$, at which a deviation from form Eq.~\ref{real} can be observed.

Based on the above discussion, we can roughly sketch the phase diagram of the different regions as shown in Fig.~5. A tough-brittle-superbrittle crossover occurs as external stress increases, while a tough-brittle crossover occurs as system size increases. Though the crossover is gradual, each region is characterized by a different form of failure probability with its own physical interpretations. The tough region has a Weibull type failure probability, and in this region, the load-sharing rule seems to be not very important. Indeed, in 2D, Curtin~\cite{c} has approximated the power-law load-sharing rule with a mean-field, global, load-sharing model. The local stress enhancement for this case is quite weak, especially in higher dimensions, and the bonds fracture relatively "independently". For the brittle region, where the "modified-Gumbel" form applies, the subsystems of size $n_c$ are such that the fracture of any single subsystem results in the global system failure. In the super-brittle region, the subsystem size $n_c$ further decreases to the order of the lattice size due to the external stress, and the strength of the system is that of the weakest bond which is distributed with the ultimate form of Eq.~\ref{real}.
\begin{figure}
  \begin{center}
    \epsfig{file=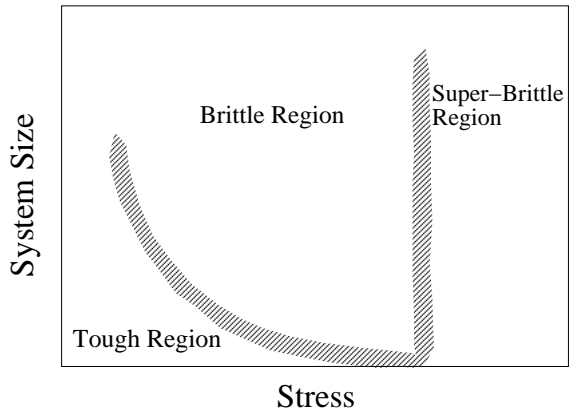,width=7.5cm,angle=0,clip=}
  \end{center}
\caption {Different regions of fracture behavior.}
\end{figure}

Given the strength distribution (Eq.~\ref{mg}), the system-size dependences of the average breakdown stress $\langle\sigma_b\rangle$ and its variance are of great interest. From Eq.~\ref{mg}, by neglecting the slow-varying factor $\log(\sigma)$ and taking the median as the average, we have
\begin{equation} 
\langle\sigma_b\rangle \sim (\log n)^{-1/\alpha} \ . 
\end{equation}
The variance can be found by considering the change in $F_n(\sigma)$ by a given factor, by which we obtain
\begin{equation}
\sqrt{\langle\sigma_b^2\rangle -\langle\sigma_b\rangle ^2} \sim (\log n)^{-(1+1/\alpha)} \ . 
\end{equation}
Thus both the strength average and its deviation decay logarithmically with increasing system size. 

\section{Other Boundary Conditions}

In the previous section, we discussed the strength distributions of heterogeneous 1D fiber-bundles with {\em interior} boundary conditions, by which each sample (considered embedded in a larger one) has both its ends held by intact neighboring bonds. The interior boundary condition gives no surface effects which, practically, are of great interest to studies of the strengths of realistic systems. Does the presence of a surface affect the strength distributions significantly? The answer is yes, and this issue has been addressed approximately by Chen and Leath~\cite{cl}, by using a transfer-matrix method with a "no-lone-bond" approximation. In this section, we shall confirm the results by an exact recursive method to calculate the failure probabilities under varying boundary conditions, namely, the {\em periodic}, {\em semi-open}, and {\em open} boundary conditions. All the boundary conditions concern the load-sharing rules at surfaces: interior boundary conditions (b.c.) promise extra unbreakable bonds outside the sample to bear stresses; open b.c. are for isolated samples with both ends open; semi-open b.c. have one end of the sample open and the other interior; periodic b.c. roll up the sample, making a closed circular system with no surface at all.

The configuration of a system sized $n$ must be one of the following:

\end{multicols}
\widetext

\begin{eqnarray*}
&(1)& \makebox[3mm]{}\underbrace{\overbrace{0\cdots0}^k1\overbrace{0\cdots 0}^l}_n \ , \qquad  k+l=n-1 \qquad \mbox{(single bond)} \ , \\
&(2)& \makebox[3mm]{}\underbrace{\overbrace{0\cdots0}^k10\cdots01\overbrace{0\cdots 0}^l}_n\ , \qquad k+l\le n-2 \qquad \mbox{(double bonds)} \ , \\
&(3)& \makebox[3mm]{}\underbrace{\overbrace{0\cdots0}^k1\overbrace{0\cdots0}^r10\cdots01\overbrace{0\cdots0}^t1\overbrace{0\cdots 0}^l}_n\ , \qquad k+r+t+l\le n-3 \qquad \mbox{(3 or more bonds)}\ , 
\end{eqnarray*}

\begin{multicols}{2}
\noindent
where $k=0,1,\cdots,n-1$ and $l=0,1,\cdots,n-k-1$ with $n=1,2,\cdots$. We define $W_k(\sigma)$ to be the survival probability of a bond with load $(1+k/2)\sigma$. Also we define $S_{n,k,l}^{(\cdot)}(\sigma)$ to be the survival probability of a system with a string of $k$ and $l$ broken bonds on each end respectively, and $F_n^{(\cdot)}(\sigma)$ to be the failure probability of the system of size $n$, under certain boundary conditions. The idea to obtain the recursion relations is the following: in general, we cut off the $0\cdots 01$ and $10\cdots 0$ strings (each has only one intact bond) from both ends and are left with a truncated sample with interior b.c. whose survival probability is available. The survival probabilities of these cut-off parts can be obtained by taking into account the specific boundary conditions which involve $F^{(i)}$ or $F^{(s)}$ for smaller systems. In doing this, one must be careful about the load re-assignments on bonds close to the surfaces. Special cases, {\em i.e.}, the single- and double-bond systems must be considered separately. Finally the recursion relations for each kind of boundary conditions turn out to be as follows:

\end{multicols}
\widetext

Interior:
\begin{eqnarray}
S_{n,k,l}^{(i)}&=&F_k^{(i)}W_{k+l}F_l^{(i)}\delta_{k+l,n-1}+F_k^{(i)}W_{n-l-2}F_{n-k-l-2}^{(i)}W_{n-k-2}F_l^{(i)} \nonumber \\ 
               & &+ {\displaystyle \sum_{r=0}^{n-k-l-3}\sum_{t=0}^{n-k-l-r-3}F_k^{(i)}W_{k+r}S_{n-k-l-2,r,t}^{(i)}W_{t+l}F_l^{(i)}} \ ,  
\end{eqnarray}

Periodic:
\begin{eqnarray}
S_{n,k,l}^{(p)}&=&F_{k+l}^{(i)}W_{2k+2l}\delta_{k+l,n-1}+F_{k+l}^{(i)}W_{n-2}F_{n-k-l-2}^{(i)}W_{n-2} \nonumber \\ 
               & &+ {\displaystyle \sum_{r=0}^{n-k-l-3}\sum_{t=0}^{n-k-l-r-3}F_{k+l}^{(i)}W_{k+r+l}S_{n-k-l-2,r,t}^{(i)}W_{k+t+l}} \ ,  
\end{eqnarray}

Semi-open:
\begin{eqnarray}
S_{n,k,l}^{(s)}&=&F_k^{(i)}W_{k+2l}F_l^{(s)}\delta_{k+l,n-1}+F_k^{(i)}W_{n-l-2}F_{n-k-l-2}^{(i)}W_{n-k+l-2}F_l^{(s)} \nonumber \\ 
               & &+ {\displaystyle \sum_{r=0}^{n-k-l-3}\sum_{t=0}^{n-k-l-r-3}F_k^{(i)}W_{k+r}S_{n-k-l-2,r,t}^{(i)}W_{t+2l}F_l^{(s)}} \ ,  
\end{eqnarray}

Open:
\begin{eqnarray}
S_{n,k,l}^{(o)}&=&F_k^{(s)}W_{2k+2l}F_l^{(s)}\delta_{k+l,n-1}+F_k^{(s)}W_{n+k-l-2}F_{n-k-l-2}^{(i)}W_{n-k+l-2}F_l^{(s)} \nonumber \\ 
               & &+ {\displaystyle \sum_{r=0}^{n-k-l-3}\sum_{t=0}^{n-k-l-r-3}F_k^{(s)}W_{2k+r}S_{n-k-l-2,r,t}^{(i)}W_{t+2l}F_l^{(s)}} \ ,  
\end{eqnarray}
\begin{multicols}{2}
\narrowtext
\noindent
with the initial condition $F_0^{(\cdot)}=1$. Thus the survival and failure probabilities of the system are given by
\begin{equation}
S_n^{(\cdot)}=\sum_{k=0}^{n-1}\sum_{l=0}^{n-k-1}S_{n,k,l}^{(\cdot)} \ ,  
\end{equation}
and
\begin{equation}
F_n^{(\cdot)}=1-S_n^{(\cdot)} \ . 
\end{equation}

These recursive relations are inter-connected with each other since $F^{(i)}$, $F^{(s)}$, $S^{(i)}$ and $S^{(s)}$ appear on the right-hand side, so they must be evaluated simultaneously. Fig.~6 Shows the exact numerical results of the fracture probabilities under various boundary conditions. It is observed that the following relation holds for most system sizes
\begin{equation}
F_n^{(o)}>(F_n^{(p)} \mbox{ and } F_n^{(s)})>F_n^{(i)} \ , 
\end{equation}
while the relation between $F_n^{(p)}$ and $F_n^{(s)}$ depends on system size and applied stress. The differences between the fracture probabilities with various boundary conditions can be orders of magnitude, especially around the optimal system size. It is thus evident that most of the failures originate from the surfaces, as stated in Ref.~\cite{cl}. We also observe a shift of the optimal system size $n_{min}^{(\cdot)}$ to a smaller value with more boundary influence. However, as system size $n$ becomes very large or the stress becomes sufficiently strong, these four probability lines merge to one, showing an identical asymptotic behavior. This means in practice, when $n$ is extremely large, the surface effects can be limited, and the interior boundary conditions can be used, and the strength distributions with all the boundary conditions will have the same general form.
\begin{figure}
  \begin{center}
    \epsfig{file=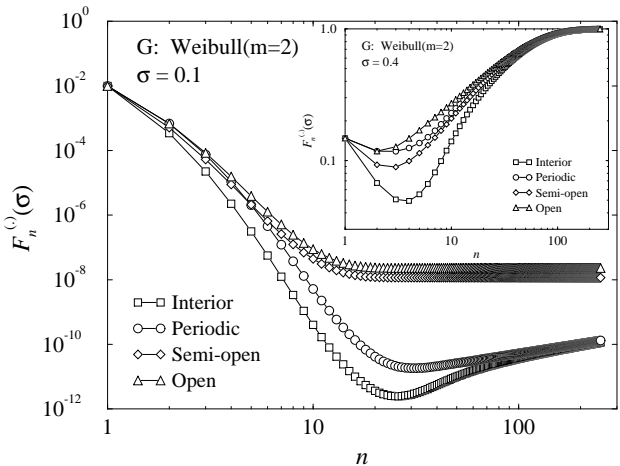,width=6.1cm,angle=-90,clip=}
  \end{center}
\caption {Fracture probabilities of systems with different boundary conditions under weak applied stress ($\sigma=0.1$). Inset: Failure probability curves under higher applied stress ($\sigma=0.4$) showing the merging curves.}
\end{figure}

\section{Some Mathematical Remarks on the Strength Distributions}

Finally, we discuss the importance of the "modified-Gumbel" form of Eqs.~\ref{gum} and \ref{mg}. We observe that the modified-Gumbel form does not itself belong to any of the three known distribution forms of extreme statistics, but its asymptotic behavior belongs to and is approximated with one of these three extreme-value-statistic distributions. Formally, it's obvious that the modified-Gumbel form $F(\sigma)=1-\exp[-an\exp(-b/\sigma^\alpha)]$ is the cdf of the minimum of $an$ independent samples drawn from a parent distribution $G(\sigma)=\exp(-b/\sigma^\alpha)$ (known as the Frechet distribution of maxima). So we obtain that the asymptotic behavior of $F(\sigma)$ for $n$ very large is Gumbel, according to the theory of extreme-value statistics. However, because of the singularity at $\sigma \rightarrow 0$, the most interesting region of fracture, its asymptotic form, the Gumbel distribution, which does not have the singularity, loses significantly its accuracy. To see this, for simplicity, let $a$, $b$ and $\alpha$ be unity. Then from the theory of extreme-value statistics, it can be shown that the asymptotic form of
 \begin{equation}
F(\sigma)=1-\exp[-n\exp(-1/\sigma)] \ , 
\end{equation}
or the domain of attraction as $n\rightarrow \infty$ is Gumbel type, {\em i.e.}, 
\begin{equation}
F(\sigma) \rightarrow 1-\exp[-\exp(\log^2 n\cdot \sigma-\log n)] \ , \label{gumbel}
\end{equation}
which serves as approximation of $F(\sigma)=1-[1-G(\sigma)]^n$ for $n$ sufficiently large. Unfortunately, this ultimate form, Eq.~\ref{gumbel}, converges very slowly. We here define the {\em convergence rate} of the asymptotic form Eq.~\ref{gumbel} as its deviation from the exact form $F(\sigma)=1-[1-G(\sigma)]^n$. Thus it can be shown that the convergence rate is about $\sim 1/\log n$ (as a comparison, the convergence rate can be $\sim 1/n$ for a fast-converging system). This makes the ultimate Gumbel form of little practical value. One way to deal with this difficulty is to make use of a Weibull-type approximation with the Weibull modulus dependent on $n$, called the penultimate form, {\em i.e.}, 
\begin{figure}[btp]
  \begin{center}
    \epsfig{file=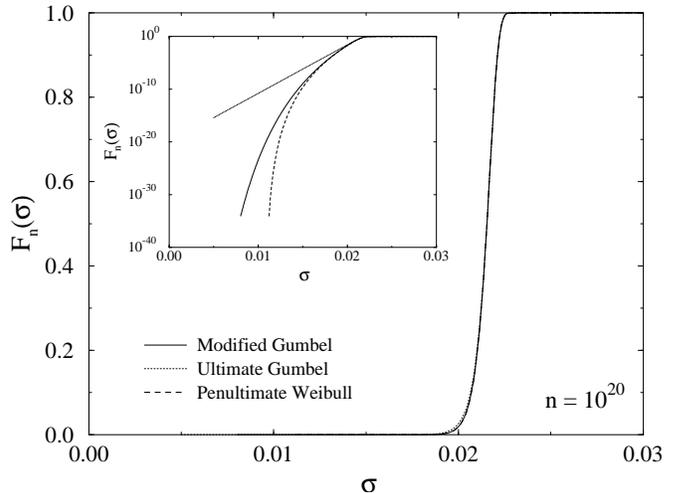,width=6.5cm,angle=-90,clip=}
  \end{center}
\caption {The modified-Gumbel distribution function $F_n(\sigma)=1-\exp[-n\exp(-1/\sigma)]$ and its ultimate asymptote, the Gumbel distribution $F_n(\sigma)=1-\exp[-\exp(\log^2n\cdot \sigma-\log n)]$ and its penultimate asymptote, the Weibull distribution $F_n(\sigma)=1-\exp[-(2\log n\cdot \sigma -1)^{\log n/2}]$, with $n=10^{20}$. Inset: A linear-log plot. Observe the differences between the distributions for $F_n(\sigma)$ small.}
\end{figure}

\begin{equation}
F(\sigma) \rightarrow 1-\exp[-(2\log n\cdot \sigma-1)^{\log n/2}] \ , 
\end{equation}
which fits very well in almost the entire range of failure probability. The penultimate Weibull form can be obtained by comparing three percentiles with those for the exact form. In this Weibull form, the modulus as a function of system size, is about $26$ for $n=10^{23}$, which is quite consistent with experiment results for the Weibull exponent in many real samples. In fact, most of the extreme-value distributions encountered in practical applications can be approximated by the penultimate Weibull form very well. This explains why the Weibull form has been so widely used in the study of breakdown phenomenon. But it's worth pointing out that the penultimate form suggests that a {\em three-parameter} fitting in the form of
\begin{equation}
F(\sigma)=1-\exp[-(a\sigma-b)^m] 
\end{equation}
should be conducted, instead of the two-parameter fitting widely used in practice (Eq.~\ref{wei}). Detailed study shows that despite the fact that the Weibull form is an excellent approximation for {\em typical average} breaking stresses, under {\em very small} stress there may still be a substantial error of up to several orders of magnitude, though the error decreases logarithmically with increasing $n$. As shown in Fig~7, the ultimate Gumbel form significantly overestimates the failure probability while the penultimate Weibull form underestimates it. In engineering applications, tests are usually done under large stresses with large failure probabilities, and are extrapolated to the weak stresses of normal operating conditions to estimate reliability. Thus this error becomes very important, especially when $n$ is not very large. The best numerical predictor of brittle failure probabilities under very weak stresses and hence most operating conditions is thus the modified-Gumbel form (Eq.~\ref{gum}) which should be used in the reliability analysis of brittle materials.

\section{Conclusion}

In summary, we have significantly simplified the recursive method for exact numerical calculations of the strength distribution of the fiber-bundle or 1D-lattice model with continuous, local, strength distribution and local load sharing. We observed qualitatively different fracture behavior for systems of different sizes and under different stresses. The critical sample size $n_c$ is found to depend upon the stress applied, a result which should generally apply in higher dimensions as well. The results here also support the conjecture that tough fracture can be described with a Weibull form and the brittle fracture with a modified-Gumbel form. Tough-to-brittle crossovers for fracture are found to occur as system sizes, or stresses change. The modified-Gumbel form can be approximated by a penultimate Weibull form for intermediate failure probabilities, but this approximation may not be applicable to very small failure probabilities due to significant errors. So the modified-Gumbel form is of special importance in the reliability analysis of brittle materials. The important effects of surfaces on the strength distributions have also been discussed with a similar exact recursion method.

\end{multicols}
\end{document}